\documentclass[preprint,12pt,3p]{elsarticle}
\usepackage{graphicx}
\usepackage{epsfig}
\usepackage{hyperref}
\usepackage{amsmath}
\usepackage{amssymb}
\usepackage{multicol}
\usepackage{float}
\usepackage{natbib}

\usepackage[T1]{fontenc}
\usepackage[document]{ragged2e}
\begin{document}   

\title{Rattling motion of proton through five membered aromatic ring systems}

\author[label1]{Somesh Chamoli\corref{cor1}}
\ead{somesh.chamoli20@gmail.com}
\author[label2]{Aniruddha Chakraborty}
\fntext[label1,label2]{School of Basic Sciences, Indian Institute of Technology Mandi, Kamand, Himachal Pradesh-175075, India.}
\begin{frontmatter}
\begin{abstract}
\noindent We study the passage of proton (H$^{+}$) through different five membered aromatic ring systems by considering one dimensional motion of the proton along a line perpendicular to the plane of the ring. The potential for the motion of (H$^{+}$) from one side of the ring to the other through the center of the ring is found to be a symmetric double well and such complexes are suggested as molecular rattles which can exhibit a ring umbrella like inversion. Our study reveals C$_{2}$H$_{2}$P$_{3}^{-}$ - H$^{+}$ to be a molecular rattle with a barrier height of 44.277 kcal mol$^{-1}$ and exciting it to the third vibrational level makes easy for the proton to go through the ring.
\end{abstract}
\date{\today}
\end{frontmatter}

\section{Introduction}
\noindent Nature contains some of the effective and the efficient molecular machines, which are present in the living system \cite{1}. The bacteria which is the smallest living body also contains molecular motors. These biological molecular machines use chemical energy in the form of ATP. These molecular motors and devices are 'molecular gears' based on triptycene molecule \cite{iwamura1988stereochemical}, 'molecular-turnstile' consisting of hexa(phenylacetylene) macrocyclic frame with a diethynlarene bridge \cite{bedard1995design}, 'molecular brakes' that is made up of a triptycene unit bonded to a bipyridine unit \cite{kelly1994molecular} and 'molecular ratchets' that is built by a triptycene molecule bonded to helicene units. Molecular machines are also based on transition metal containing rotaxanes and catenanes. In catenanes one ring can slide within another ring and rotaxanes have been used as molecular frameworks for the preparation of switchable, bistable systems \cite{anelli1991molecular}. A light powered molecular motor is also created by koumura and coworkers \cite{koumura1999light}. Among all these molecular machines there exist a class of molecules and their complexes with ions named as molecular rattle which can exhibit an umbrella like inversion exactly as ammonia molecule. The interaction of ions with the ring systems is origin for a variety of applications in gas sensors \cite{jiang2009porous}, hydrogen storage \cite{cabria2005enhancement}, isotopic separation \cite{schrier2012thermally}, lithium ion batteries \cite{yao2012diffusion} and desalination of water \cite{cohen2012water}.
\subsection{Possibility}
In electrophilic aromatic substitution two types of intermediates are formed : $ \sigma $-complex and $ \pi $-complex. $ \pi $-complex is formed first before the formation of $ \sigma $-complex  and then is converted to $ \sigma $-complex. This $ \pi $-complex is just a non- covalent molecular interaction between ion and the $ \pi $ electron cloud of the aromatic ring system. In a $ \pi $-complex an ion(proton) can reside either above or below the plane of the aromatic ring system as shown in Fig. 1 , so there arises a possibility that this proton could oscillate between these two positions in a $ \pi $-complex  through the centroid of the ring.

\begin{figure}
    \centering
    \includegraphics{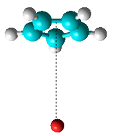}
    \includegraphics{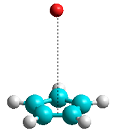}
    \caption{ $ \pi $-complex of proton with cyclopentadienyl anion.}
    \label{fig:my_label1}
\end{figure}
The potential for the motion of proton from one side of the ring to the other side is a double well and the minima in the double well correspond to the stable geometries of the proton on either side of the ring, and the transition state corresponds to the geometry in which the proton is in plane with the ring at its center as shown in the Fig. 2.

 \begin{figure}[h!]
     \centering
     \includegraphics[scale=0.65]{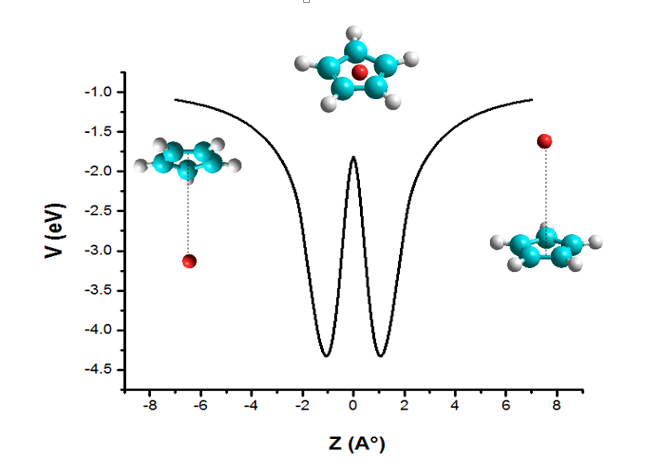}
     \caption{The double well potential curve}
     \label{fig:my_label2}
 \end{figure}
Such a possibility of rattling motion of proton through the benzene system is already examined by N. Sathyamurthy {\it et. al.}, \cite{shresth1996possibility}. They monitored the interaction energy for benzene-proton system and the resulting potential energy curve was double well with the minima at $\pm$ 1.1  {\AA}  with the barrier height that proton was facing was 59.5 kcal/mol[LCAO-MO-SCF approach with 6-311G basis set]. In a related paper, the possibility of lithium ion through cyclononatetraenyl (CNT) anion was studied and the barrier height for that motion was found to be 11.50 kcal/mol \cite{das2002through}. Rattling motion of different alkali metal ions through the cavities of model compounds of graphyne and graphdiyne is also studied and the motion of $ Li^{+} $ across the rings of C$_{12}$H$_{6}$, C$_{24}$H$_{12}$, C$_{26}$H$_{12}$, C$_{28}$H$_{12}$ and C$_{30}$H$_{12}$ occurs with the energy barriers in the range of 0-5 kcal/mol \cite{chandra2013rattling}.
\section{Method} 
\label{appendix-sec1}
All the single point energy calculations and geometry optimization mentioned in this paper are done using ab initio (6-31G** and 6-311G** basis set ) method. The ab initio calculations are done using hyperchem package \cite{release20027}. Initially the geometry of different aromatic  rings was optimized and the energy after optimization ({\it i.e.}, $ E_{ring} $ ) was calculated. The energy of the proton ({\it i.e.}, $ E_{proton} $ ) was also calculated. Now taking the optimized structures of the rings and placing the proton at various positions (Z) along an axis which is perpendicular to the ring plane and passing through the centroid of the ring system, the energy of complex with proton for different rings ({\it i.e.}, $ E_{complex} $ ) was calculated. For each position of proton a single point energy calculation was done. Using these values of $ E_{complex} $, interaction energy (V) between proton and the ring system can be evaluated as :                
\begin{equation}\label{eq:1}
V =  E_{complex} - E_{ring} - E_{proton}
\end{equation}                       
Corresponding to each and every position of proton, interaction energy was calculated and from the interaction energy data for different position we were able to find the minimum value of interaction energy({\it i.e.}, minima ) as well as interaction energy for the transition state {\it i.e.}, interaction energy when the proton is exactly at the centre of aromatic ring as shown in Fig. 3.

\begin{figure}[h!]
    \centering
    \includegraphics[scale=1]{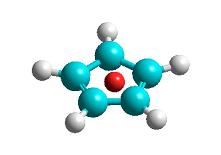}\\
 \caption{ Transition state.}
    \label{fig:my_label3}
\end{figure}
With the help of interaction energy for transition state and minimum value of potential, energy barriers for the motion of proton through the ring is calculated :
      \begin{equation}\label{eq:2}
E_{barrier} = V_{transition~state}-  V_{minimum}
\end{equation}    
\section{Results and Discussion}
Initially we took a five membered aromatic ring which is cyclopentadienyl anion (C$_{5}$H$_{5}^{-}$) and optimize its geometry and did single point energy calculations for each and every location of proton.
\begin{figure}[h!]
    \centering
 \includegraphics[scale=0.7]{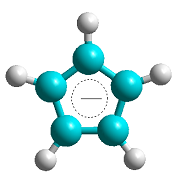}\\

\caption{ cyclopentadienyl anion.} 
    \label{fig:my_label4}
\end{figure}
We use 6-311 G ** basis set for optimizing as well as for single point energy calculations. Fig. 4 shows the optimized geometry of the cyclopentadienyl anion and Table 1 shows the bond distances of optimized structure of cyclopentadienyl anion.
\begin{figure}[h!]
    \centering
\includegraphics[scale=0.60]{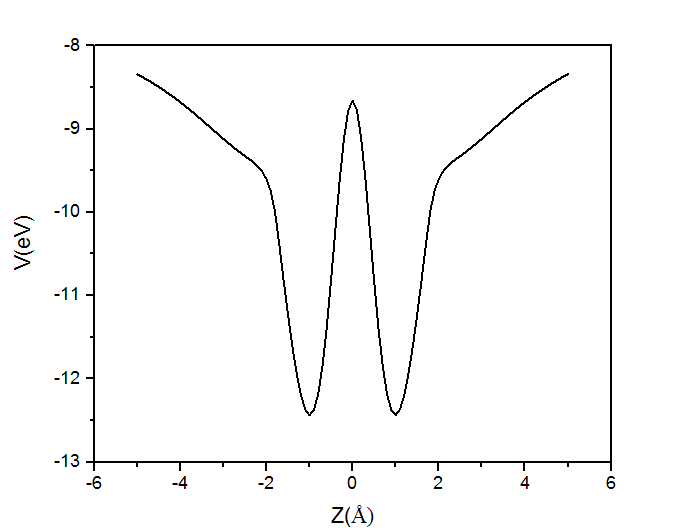}
\caption{Interaction potential for C$_{5}$H$_{5}^{-}$-H$^{+}$.}
\label{fig:my_label5}
\end{figure}
\begin{table}[h!]
\Centering
    \begin{tabular}{||c|c|| }
    \hline
    C-C bond length &C-H bond length\\
    \hline
     1.403 \AA      & 1.079 \AA\\ 
     \hline
    \end{tabular}
\caption{ Bond distances of C$_{5}$H$_{5}^{-}$.}
\end{table}
\noindent
From Fig. 5 we can see the interaction potential energy curve, which gives us the interaction minima at $\pm$ 1.0 \AA~and a barrier height of 87.101 kcal/mol. As compared to the benzene-proton system the barrier height which proton is facing while passing through this cyclopentadienyl anion ring is more, this is probably due to the pore size of the ring, as cyclopentadienyl anion is five membered ring while benzene is six membered ring due to which motion of proton becomes more difficult.
\noindent
Since proton is facing more barrier so we decided to make some substitution in the ring so that this barrier can be reduced and proton can easily pass through the aromatic ring. The first substitution we did is replacing two carbon atoms with two nitrogen atom making the ring as 2H-imidazol-2-ide commonly known as imidazolate anion. Using 6-31G** basis set (LCAO-MO SCF) we optimized the structure. Fig. 6 shows the optimized geometry of the imidazolate anion and the details of the optimized structure are given in the Table 2.
\begin{figure}[h!]
    \centering
    \includegraphics[scale=0.48]{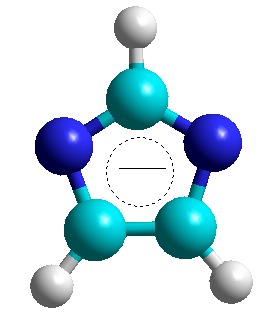}
    \caption{imidazolate anion.}
    \label{fig:my_label6}
\end{figure}
\begin{table}[h!]
\Centering
    \begin{tabular}{||c|c|c|c|| }
    \hline
    N$_{1}$-C$_{2}$ bond length &N$_{1}$-C$_{5}$ bond length &C$_{4}$-C$_{5}$ bond length &C-H bond length\\

    \hline
     1.327 \AA      & 1.357 \AA    & 1.078 \AA   & 1.372 \AA  \\ 
     \hline
    \end{tabular}
\caption{bond distances of C$_{3}$H$_{3}$N$_{2}^
{-}$.}
\end{table}  
\begin{figure}[h!]
    \centering
    \includegraphics[scale=0.60]{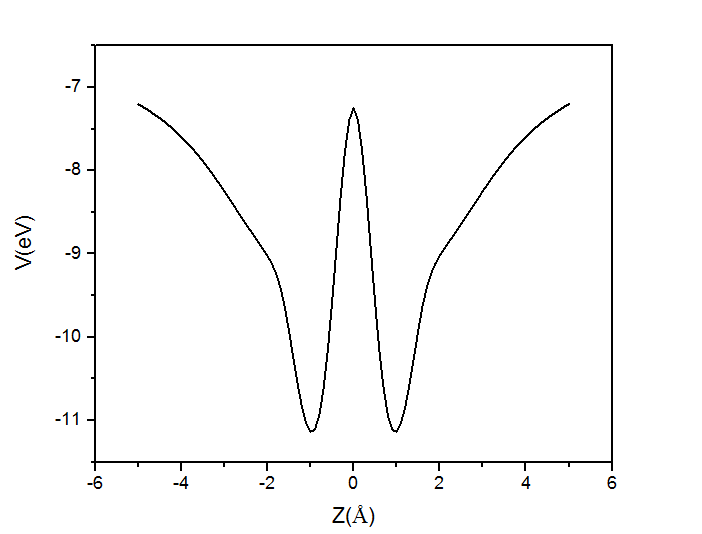}
    \caption{Interaction potential for C$_{3}$H$_{3}$N$_{2}
^{-}$-H$^{+}$.}
    \label{fig:my_label7}
\end{figure}
Fig. 7 gives us interaction minima at $\pm$ 1.0 \AA~and a barrier potential equals to 89.599 kcal/mol which was 2.498 units more than the barrier potential for cyclopentadienyl anion ring. 
So on comparing results of barrier height from cyclopentadienyl anion ring, and imidazolate anion ring we conclude that since imidazolate anion ring contains 2 nitrogen atom and hence its effective pore size will be smaller than cyclopentadienyl anion ring that's why proton is facing more barrier while passing through imidazolate anion ring. Now we want to check that what happens on substitution of 3 carbon atoms with nitrogen atoms, so we optimized the geometry of 1,2,3-triazol-2-ide anion using 6-31G** basis set still using ab-initio[LCAO-MO-SCF] method. Fig. 8 represents optimized geometry of 1,2,3-triazolate anion ring and the bond lengths of this optimized structure are given in Table 3. Fig. 9 shows the interaction minima at

 \begin{figure}[h!]
    \centering
    \includegraphics[scale=0.5]{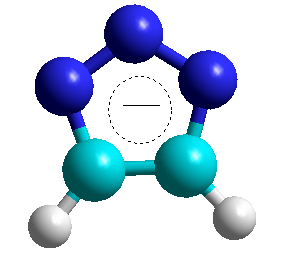}
\caption{1,2,3-triazol-2-ide anion}
    \label{fig:my_label8}
\end{figure}
\begin{table}[h!]
\centering
    \begin{tabular}{||c|c|c|c|| }
    \hline
    N-N bond length&N-C bond length&C-H bond length& C-C bond length\\

    \hline
     1.308 \AA      & 1.334 \AA    & 1.074 \AA    & 1.376 \AA\\ 
     \hline
    \end{tabular}
\caption{ Bond distances of C$_{2}$H$_{2}$N$_{3}^{-} $.}
\end{table}
\begin{figure}[h!]
    \centering
    \includegraphics[scale=0.6]{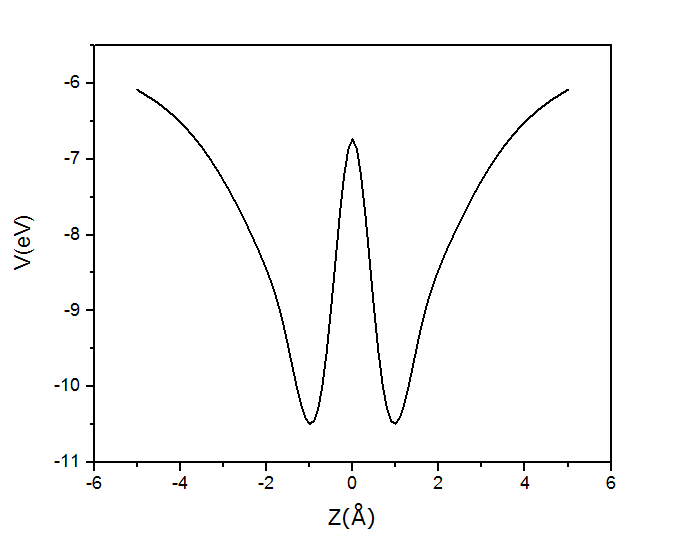}
\caption{Interaction potential for $ C_{2}H_{2}N_3^{-}$-H$^{+}$.}
    \label{fig:my_label9}
\end{figure}
$\pm$ 1.0 \AA~ and gives us a barrier height of 86.709 kcal/mol which is 2.890 units less than the barrier height for the imidazolate anion ring. This may be due to the 3 more electronegative nitrogen atom which perturbs the electronic cloud somehow not completely but up to little extent may be and due to which the ring becomes deshielded for the proton passage. Since proton is facing more barrier so we decided to make some more substitution in the ring so that this barrier can be reduced and proton can easily pass through the aromatic ring. The next substitution we did is replacing one carbon atom having negative charge with the phosphorous atom making the ring as phosphacyclo-pentadienyls commonly known as Phospholide anion.
Using 6-311G** basis set (LCAO-MO SCF) we optimized the structure. Fig. 10 shows the optimized Geometry of the Phosphol-1-ide anion and the details of the optimized structure are given in the Table 4.
\begin{figure}[h!]
    \centering
    \includegraphics[scale=0.8]{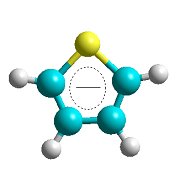}
    \caption{phosphol-1-ide anion.}
    \label{fig:my_label10}
\end{figure}
\begin{table}[h!]
\small
\begin{tabular}{||c|c|c|c|c|| }
    \hline
   P-C$_{1} $ bond length& C$_{1}$-H bond length& C$_{1}$-C$_{2}$ bond length& C$_{2}$-H bond length& C$_{2}$-C$_{3}$ bond length \\

    \hline
1.764 \AA & 1.078 \AA    & 1.376 \AA    & 1.080 \AA    & 1.416 \AA\\ 
     \hline
    \end{tabular}
\caption{ Bond distances of C$_{4}$H$_{4}$P$^{-}$.}
\end{table}
\begin{figure}[h!]
    \centering
    \includegraphics[scale=0.6]{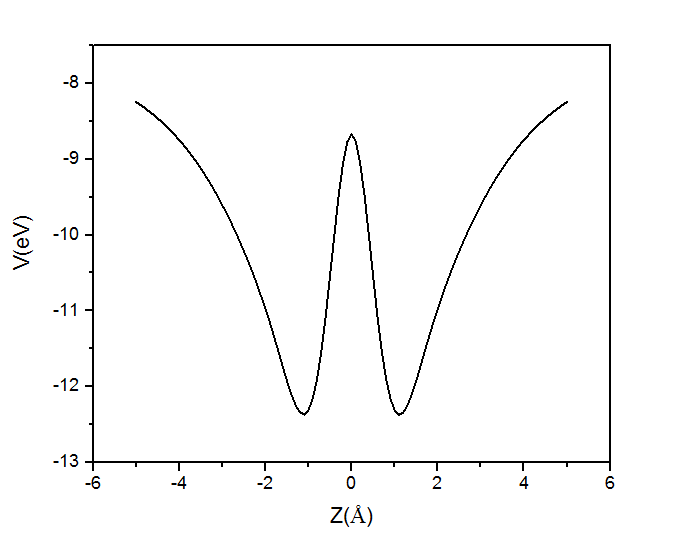}
    \caption{Interaction potential for C$_{4}$H$_{4}$P$^{-}$-H$^{+}$ .}
    \label{fig:my_label11}
\end{figure}
\noindent From Fig. 11 we can see that the substitution of phosphorus atom in the cyclopentadienyl anion ring changes the position of minima from $\pm$ 1.0 \AA~ to $\pm$ 1.1 \AA~ that is slightly changing the width of barrier. The barrier height that proton is facing while passing through Phospholide anion ring is 85.417 kcal/mol, which is lower than 87.101 kcal/mol by 1.684 units only. Keeping in mind that one additional nitrogen atom reduces the barrier for proton passage through imidazolate anion by 2.890 kcal/mol and substituting one carbon atom of cyclopentadienyl anion by phosphorous atom reduces barrier potential by 1.684 units. We substitute
\begin{figure}[h!]
    \centering
    \includegraphics[scale=0.45]{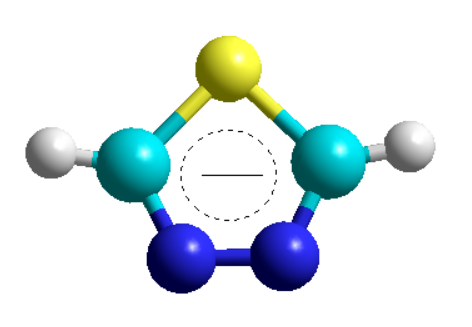}
    \caption{1,2,4-diazaphosphol-4-ide anion.}
    \label{fig:my_label12}
\end{figure}
\begin{table}[h!]
\Centering
    \begin{tabular}{||c|c|c|c|| }
    \hline
    P-C bond length & C-N bond length &N-N bond length &C-H bond length\\

    \hline
     1.760 \AA    &1.303 \AA    &1.348 \AA   &1.079 \AA\\ 
     \hline
    \end{tabular}
\caption{bond distances of C$_{2}$H$_{2}$PN$_{2}^{-}$.}
\end{table}
carbon atoms of cyclopentadienyl anion by one phosphorus and two nitrogen atom resulting in 1,2,4-diazaphosphol-4-ide. 
\begin{figure}[h!]
    \centering
    \includegraphics[scale=0.6]{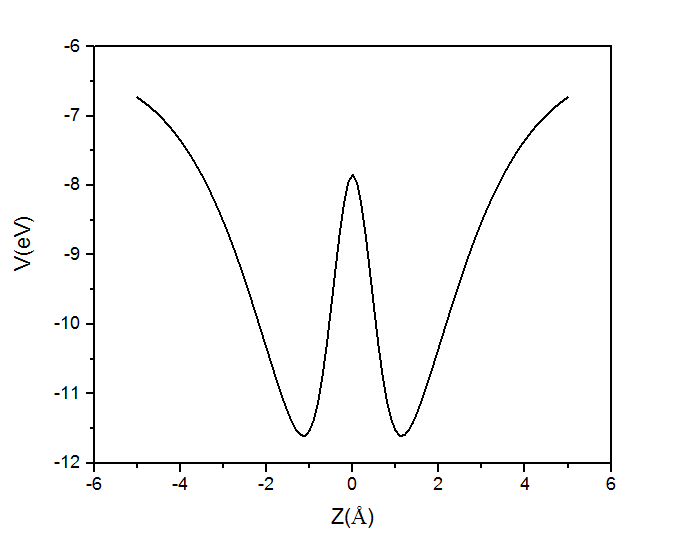}
    \caption{Interaction potential for C$_{2}$H$_{2}$PN$_{2}^{-}$
 - H$^{+}$.}
    \label{fig:my_label13}
\end{figure}
Using 6-31G ** basis set we optimize its structure. Fig. 12 shows the optimized structure of 1,2,4-diazaphosphol-4-ide and structural details of optimized structure are given in Table 5. Fig. 13 shows interaction potential energy curve for 1,2,4-diazaphosphol-4-ide anion and proton, which shows interaction minima at $\pm$ 1.1 \AA ~ and gives a barrier height of 86.780 kcal/mol which was closer to 1,2,3 triazolate anion and proton system. Therefore the attempt of lowering the barrier height by using one phosphorus and two nitrogen atoms does no such effect on barrier potential that proton was facing while passing through the ring. Now in order to lower the barrier less than 86.780 kcal/mol we substitute all carbon atoms of cyclopentadienyl anion ring completely by three phosphorus and two nitrogen atoms making it 1,3,2,4,5-diazatriphosphol-2-ide. Fig. 14 shows the optimized geometry of 1,3,2,4,5-diazatriphosphol-2-ide which was optimized using 6-31G** basis set and the structural details are in the Table 6.
\begin{figure}[h!]
    \centering
    \includegraphics[scale=0.4]{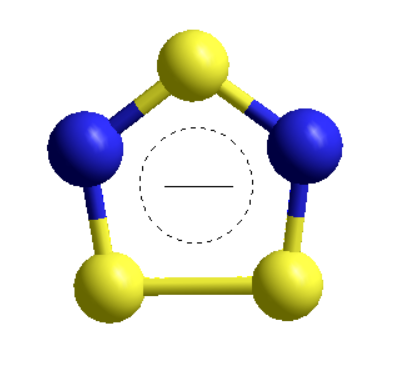}
    \caption{ 1,3,2,4,5-diazatriphosphol-2-ide anion.}
    \label{fig:my_label14}
\end{figure}
\begin{table}[h!]
\Centering
    \begin{tabular}{||c|c|c|| }
    \hline
    P-N bond length &P-P bond length &N-P bond length\\

    \hline
    1.616 \AA   & 2.091 \AA    & 1.651 \AA   \\ 
     \hline
    \end{tabular}
\caption{bond distances of P$_{3}$N$_{2}^{-}$.}
\end{table}
\begin{figure}[h!]
    \centering
    \includegraphics[scale=0.6]{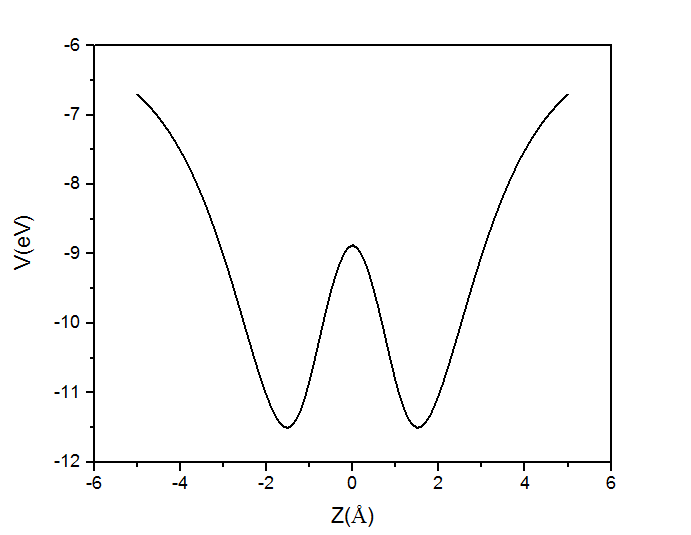}
    \caption{Interaction potential for P$_{3}$N$_{2}^{-}$}
    \label{fig:my_label15}
\end{figure}
Fig. 15 shows interaction potential energy curve for 1,3,2,4,5-diazatriphosphol-2-ide anion and proton system and gives an interaction minimum at $\pm$ 1.5 \AA~ and a barrier height of 60.649 kcal/mol which was 26.452 kcal/mol less than the barrier height for cyclopentadienyl anion and proton system.  So this last substitution decreases the barrier height for proton passage but increases the barrier width as its interaction minimum is at $\pm$ 1.5 \AA~ whereas for cyclopentadienyl anion ring it was $\pm$ 1.0 \AA. Therefore this substitution can be considered well according to barrier height but not according to barrier width. Since one phosphorus was lowering the barrier height by $1.684$ units so we took three phosphorus atoms because the pore size of nitrogen is smaller than carbon and also its addition does not affect the barrier height much so finally we substitute only three carbon atoms of cyclopentadienyl anion with three phosphorus atoms and the rest of the atoms are not affected. The resulting structure becomes 1,2,4-triphosphol-4-ide anion.

\begin{figure}[h!]
    \centering
    \includegraphics[scale=0.6]{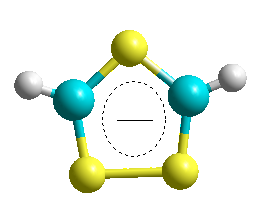}
    \caption{1,2,4-triphosphol-4-ide anion.}
    \label{fig:my_label16}
\end{figure}
\begin{table}[h!]
\Centering
\begin{tabular}{||c|c|c|| }
    \hline
    C-H bond length &P-P bond length &C-P bond length\\
    \hline
     1.079 \AA      & 2.116 \AA    & 1.726 \AA   \\ 
     \hline
\end{tabular}
\caption{bond distances of C$_{2}$H$_{2}$P$_{3}^{-}$.}
\end{table}
\begin{figure}[h!]
    \centering
    \includegraphics[scale=0.6]{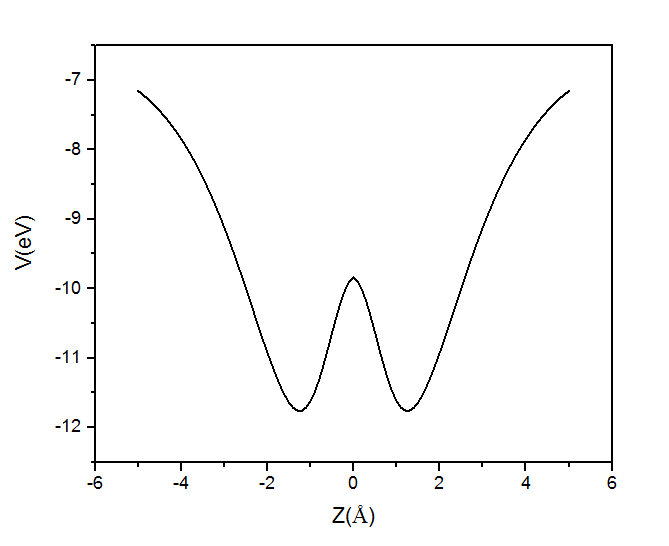}
    \caption{Interaction potential for C$_{2}$H$_{2}$P$_{3}^
{-}$ - H$^{+}$.}
    \label{fig:my_label17}
\end{figure}
 Using 6-31 G ** basis set we optimized structure of 1,2,4-triphosphol-4-ide anion. Fig. 16 shows the optimized structure of 1,2,4-triphosphol-4-ide and the structural details of optimized structure are given in Table 7. Fig. 17 shows interaction potential energy curve for 1,2,4-phosphol-4-ide anion and proton, which shows the interaction minima at $\pm$ 1.2 \AA , and gives a barrier potential of 44.277 kcal/mol which is almost half of the barrier potential for cyclopentadienyl anion ring, so substituting three carbon atoms with phosphorous atoms increases width of barrier potential by $0.1$ units but decreases the barrier height by $42.824$ units making it more easier for the proton to pass through the ring.
After finding that motion of H$^{+}$ through 1,2,4-triphosphol-4-ide anion ring is facile as proton is not facing too much barrier height, we calculate the vibrational energy levels. Since, the potential for the proton passage was a symmetric double well, so we model it by an analytical form of a quartic double well which is given by: 
\begin{equation}\label{eq:3}
V(z) = a + bz^{2}+cz^{4}
\end{equation}
where a = -10.1947, b = -1.55617 and c = 0.364613 are optimized values obtained by fitting the interaction potential data, which is done using Wolfram's Mathematica\cite{mathematica}. We create a data table for computed potential and also for the analytical potential given by Eq. \ref{eq:1}. Defining a new function g[a,b,c] as given below
 \begin{equation}\label{eq:4}
g[a,b,c] = \sum_{i=1}^{n} [V(i)_{computed}- V(i)_{analytical}]^2/n
\end{equation}
and obtaining minimum of g[a,b,c] gives us the optimized value for the unknown parameters a, b and c.
\begin{figure}[h!]
    \centering
    \includegraphics[scale=0.9]{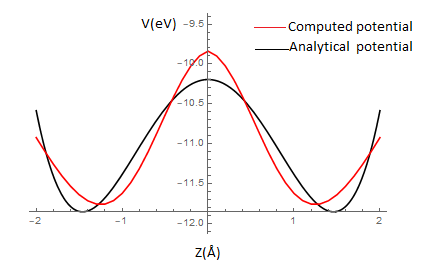}
    \caption{Comparison of the computed and analytical potential.}
    \label{fig:my_label18}
\end{figure}
\begin{figure}[h!]
    \centering
    \includegraphics[scale=0.6]{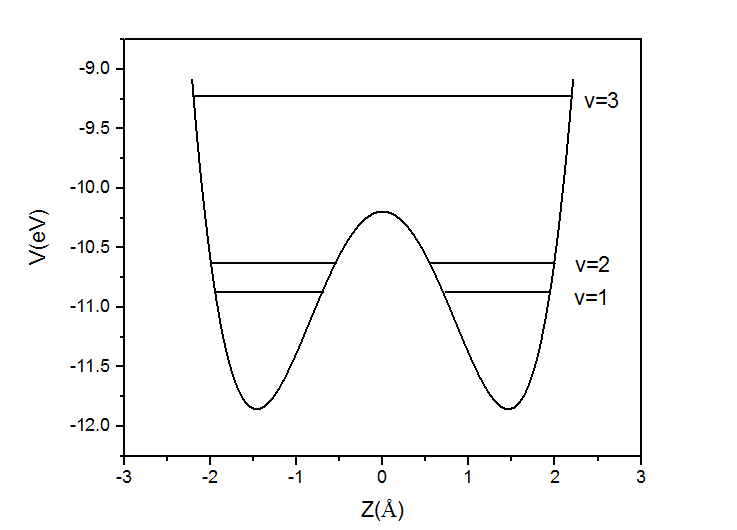}
    \caption{Vibrational energy levels for proton oscillation in C$_{2}$H$_{2}$P$_{3}^{-}$ - H$^{+}$ complex.}
    \label{fig:my_label19}
\end{figure}
Fig. 18 shows a comparison between computed double well potential and the analytical potential of eq \ref{eq:3}.
Using the above potential, Time independent Schrodinger equation for one dimensional motion of proton along $z$ axis is also solved using Mathematica which gives us vibrational energy quanta. The Hamiltonian matrix was formed using Harmonic oscillator basis and a $60\times60$  hamiltonian matrix was then diagonalized to obtain the converged lowest five vibrational energy quanta. The resultant few vibrational energy levels are shown in Fig. 19. From Fig. 19 we can see that for E > E$_{barrier}$ the proton shows free oscillation between the two sides of the 1,2,4-triphosphol-4-ide anion ring, while for E <  E$_{barrier}$ , we may find the possibility of quantum tunneling. For E <<  E$_{barrier}$, the proton oscillate around the minima on either side of the ring. Thus exciting the motion of proton to third vibrational level (v=3), it is possible for the proton to show rattling motion through 1,2,4-triphosphol-4-ide anion ring as the third vibrational level is above the height of the energy barrier.
\section{Conclusions}
\noindent We have studied the possibility of proton passing through different five membered aromatic rings by  allowing it to move through the center of the ring along a straight line perpendicular to the plane of the ring and we found that the interaction potential is represented by a double well. We took cyclopentadienyl anion ring as a parent ring and we do different substitution in the ring to make its barrier height lower. The barrier height for the proton motion through these different substituted rings lies in the range of 44-90 kcal/mol. We found that the proton passage is easier and it is possible for the proton to pass through the ring when it is substituted by three phosphorus atoms making it 1,2,4-triphosphol-4-ide, which is also aromatic and shows an interaction minima at 1.2 \AA~on either side of the ring and gives a barrier height of only 44.277 kcal/mol. Since the barrier height associated with C$_{2}$H$_{2}$P$_{3}^{-}$ - H$^{+}$ is not high, it becomes possible to perceive the oscillating behavior of proton. Our study speculate that excitation to third vibrational level leads to proton-oscillation in which proton goes from one side to other side of 1,2,4-triphosphol-4-ide anion ring.
\section*{Acknowledgments}
The authors thank Dr. Bidisa Das for useful discussions on various aspects of the problem.

\end{document}